\documentclass[acmsmall,screen]{acmart}

\usepackage{fancyhdr}
\usepackage{listings}
\AtBeginDocument{%
  }

\setcopyright{none}
\copyrightyear{2024}
\acmYear{2024}
\acmDOI{}

\acmConference[ICFP miniKanren '24]{International Conference on Functional Programming, miniKanren and Relational Programming Workshop}{Sept 06, 2024}{Milan, Italy}
\acmISBN{978-1-4503-XXXX-X/18/06}

\begin{document}

\title{To Be or Not To Be: Adding Integrity Constraints to stableKanren to Make a Decision}

\author{Xiangyu Guo}
\email{Xiangyu.Guo@asu.edu}
\orcid{0000-0001-8120-2365}
\affiliation{%
  \institution{Arizona State University}
  \streetaddress{699 S. Mil Ave}
  \city{Tempe}
  \state{Arizona}
  \country{USA}
  \postcode{85281}
}

\author{Ajay Bansal}
\email{Ajay.Bansal@asu.edu}
\orcid{0000-0001-8639-5813}
\affiliation{%
  \institution{Arizona State University}
  \streetaddress{699 S. Mil Ave}
  \city{Tempe}
  \state{Arizona}
  \country{USA}
  \postcode{85281}
}

\renewcommand{\shortauthors}{Xiangyu Guo and Ajay Bansal}

\settopmatter{printacmref=false}
\settopmatter{printfolios=true}
\renewcommand\footnotetextcopyrightpermission[1]{}
\pagestyle{fancy}
\fancyfoot{}
\fancyfoot[R]{miniKanren'24}
\fancypagestyle{firstfancy}{
  \fancyhead{}
  \fancyhead[R]{miniKanren'24}
  \fancyfoot{}
}
\makeatletter
\let\@authorsaddresses\@empty
\makeatother

\begin{abstract}
  We integrate integrity constraints to stableKanren to enable a new problem-solving paradigm in combinatorial search problems.
  stableKanren extends miniKanren to reasoning about contradictions under stable model semantics.
  However, writing programs to solve combinatorial search problems in stableKanren did not fully utilize the contradiction reasoning.
  This is mainly due to the lack of control over the predicate (goal function) outcome during resolution.
  Integrity constraints defined by answer set programming (ASP) provide the ability to constrain the predicate outcome.
  However, integrity constraints are headless normal clauses, and stableKanren cannot create a goal function without a valid head.
  There are two approaches to handling integrity constraints, but they do not fit stableKanren.
  Therefore, we design a new approach to integrate integrity constraints into stableKanren.
  We show a uniform framework to solve combinatorial search problems using integrity constraints in extended stableKanren.
\end{abstract}



\received{10 June 2024}
\received[revised]{5 July 2024}
\received[accepted]{28 August 2024}

\maketitle
\thispagestyle{firstfancy}

\section{Introduction}
    Daniel Friedman et al. built miniKanren to capture the essence of Prolog and show a natural way to extend functional programming to relational programming \cite{friedman2005reasoned}.
    The core miniKanren implementation introduces only a few operators to users: \textit{==} for \emph{unification}, \textit{fresh} for \emph{existential quantification}, \textit{conde} for \emph{disjunction}, and a \textit{run} interface.
    The \emph{conjunction} is captured by the sequential evaluation process naturally, so there is no operator for conjunction.
    
    The simple design and convenient access to the underlying resolution and unification make extending miniKanren easy.
    Xiangyu Guo et al. extended miniKanren under stable model semantics (Definition \ref{def:stable-model}) to support reasoning about contradiction in stableKanren \cite{stableKanren}.
    For example, Alice and Bob are planning a business trip, but only one person can travel, either Alice or Bob.
    The condition can be represented as a stableKanren program as follows.
    \begin{lstlisting}
(defineo (Alice)      (defineo (Bob)
  (noto (Bob))          (noto (Alice)))
    \end{lstlisting}
    In addition to miniKanren's operators, stableKanren introduces two new operators, \textit{defineo} and \textit{noto}.
    The \textit{noto} operator represents the negation (\textit{not}) in the normal program clause (Definition \ref{def:normal-program-clause}).
    Also, stableKanren uses \textit{defineo} to define its goal functions instead of \textit{define} in miniKanren.
    In our example, no variable exists in each goal function, so it is an equivalent of propositional logic.
    The success of the goal function means the person can travel, and failure means they cannot.
    The same \textit{run} interface is used for the query in stableKanren.
    A set of queries and outputs is shown as follows.
    \begin{lstlisting} 
> (run 1 (q) (Bob))
(*\textbf{(\_.0)}*)
> (run 1 (q) (Alice) (Bob))
(*\textbf{()}*)
    \end{lstlisting}
    The \textit{run} interface has three parameters.
    The first parameter $n$ is the number of answers we expect; the query returns at most $n$ answers.
    The second parameter is the query variable, which stores the answers found by the query.
    The third parameter is the actual query.
    The first query asks: ``Can Bob travel?''
    It returns a list containing one element ``\_.0'', a representation of anything in miniKanren and stableKanren.
    Anything can let goal function \textit{Bob} succeed.
    Hence, Bob can travel.
    The second query asks: ``Can Alice and Bob travel together?''
    It returns an empty list, representing nothing in miniKanren and stableKanren.
    In this case, nothing can let both goal functions \textit{Alice} and \textit{Bob} succeed.
    Therefore, they cannot travel together.

    The new \textit{noto} (negation) operator in stableKanren can also be used for writing programs to solve combinatorial search problems.
    For example, the nqueens problem is given $n$ queens to place them on a $n \times n$ chessboard so that no two queens can attack one another.
    One way to write the nqueens solution in stableKanren with non-relational helpers (gt, sub, diagonal) is as follows.
    \begin{lstlisting}
  (defineo (queens x n qi qo)                           
    (conde                             (defineo (nqueens n q)
      [(== x 0) (== qi qo)]              (queens n n `() q))
      [(fresh (x1 y qn)                            
        (*\textbf{(gt x 0)}*)                         (defineo (attack x y qs)
        (pickqueen x y n)                (conde
        (noto (attack x y qi))             [(fresh (h t)
        (*\textbf{(sub x 1 x1)}*)                            (== `(,h . ,t) qs)
        (== `((,x ,y) . ,qi) qn)              (attack x y t))]
        (queens x1 n qn qo))]))            [(fresh (x1 y1 t)
                                              (== `((,x1 ,y1) . ,t) qs)
  (defineo (pickqueen x y n)                  (== x1 x))]
    (conde                                 [(fresh (x1 y1 t)
      [(*\textbf{(gt n 0)}*) (== y n)]                      (== `((,x1 ,y1) . ,t) qs)
      [(fresh (n1)                            (== y1 y))]
        (*\textbf{(gt n 1)}*)                            [(fresh (x1 y1 t)
        (*\textbf{(sub n 1 n1)}*)                            (== `((,x1 ,y1) . ,t) qs)
        (pickqueen x y n1))]))                (*\textbf{(diagonal x y x1 y1)}*))]))
    \end{lstlisting}
    The \textit{queens} directs the search process in \textit{pickqueen} and backtracks on the conflicted queens in \textit{attack}.
    Firstly, \textit{pickqueen} picks one new queen to place in an open position.
    Then, \textit{attack} checks the new queen with the existing queens on the board to ensure they are not attacking each other from row, column, and diagonal.
    Lastly, if the checking fails, the new queen is discarded.
    Otherwise,  the new queen is added to the existing queens, and the process is repeated to place the next queen.
    An answer is found when it successfully places $n$ queens.

    The above program did not fully utilize the \textit{noto} capability provided by stableKanren, that it can write programs to produce multiple models under stable model semantics (Definition \ref{def:stable-model}).
    For example, given four numbers, a program to generate all $2^{4*4}$ queen placements on the $4 \times 4$ chessboard is in Listing \ref{lst:sk_comb}.
    \begin{lstlisting} [caption=A program generate all queen placements in stableKanren, label=lst:sk_comb]
(defineo (num x) (conde [(== 1 x)] [(== 2 x)] [(== 3 x)] [(== 4 x)]))
(defineo (pick x y) (num x) (num y) (noto (free x y)))
(defineo (free x y) (num x) (num y) (noto (pick x y)))
    \end{lstlisting}
    It uses \textit{pick} and \textit{free} to create a selection saying either to pick or not to pick a queen to place in position $(x, y)$.
    Therefore, it generates all combinations, from no queens placed on the board to 16 queens placed on the board.
    Among all combinations, the answers are those combinations that place four queens on the board and do not attack each other.
    However, stableKanren cannot constrain the outcome of \textit{pick} and \textit{free}, making it impossible to select the expected answers.

    Answer set programming (ASP) defines integrity constraints (Definition \ref{def:constraint-rule}) that can control the outcome of the predicate (goal function).
    We show in Section \ref{sec:predicate-constraints} that integrity constraints are headless normal clauses, but stableKanren requires a valid head to create a goal function.
    Some underlying changes are needed to support integrity constraints in stableKanren.
    We discuss two existing approaches to handling integrity constraints in Section \ref{sec:bottom-up-top-down}, and we show that they are not suitable for stableKanren.
    Hence, we need to design a new approach to extend stableKanren with integrity constraints.

    We treat integrity constraints as resolution exceptions, and violated exceptions prune out the resolution branch.
    In Section \ref{sec:design-idea}, we demonstrate the idea of breaking an integrity constraint into emitters (Definition \ref{def:emitter}) and verifiers (Definition \ref{def:verifier}).
    The emitters are implanted as checkpoints in resolution to emit values to verifiers.
    Once verifiers collect sufficient values, they verify the constraint; the verification result controls resolution.
    In Section \ref{sec:constraint-compilation} and \ref{sec:constraint-checking}, we extend stableKanren with integrity constraints using the method we presented.
    We introduce a new macro \textit{constrainto} to compile integrity constraints as emitters and verifiers.
    Then, the emitters are implanted as checkpoints in the stableKanren's resolution.
    Lastly, we illustrate how the verifiers collect values from corresponding emitters during resolution.
    Once verifiers collect sufficient values, they verify the constraint.
    If the verifiers find a violated constraint, it terminates the resolution branch.
    After we finish these enhancements, we can write the following integrity constraint in extended stableKanren to prevent picking queens on the same row.
    \begin{lstlisting}
(constrainto [(pick x y) (pick u v)] [(= x u) (not (= y v))])
    \end{lstlisting}
    In Section \ref{sec:result}, we demonstrate using integrity constraints with three examples: the nqueens problem, graph coloring, and the Hamiltonian cycle.
    Instead of describing ``how'' to solve these problems, we can specify ``what'' properties the problem's solutions should have using integrity constraints.
    All those problems are modeled as graphs.
    Each graph generates all combinations of the nodes or edges that are picked or not picked.
    Eventually, unwanted combinations are pruned using integrity constraints derived from the desired solution's property.
    In Section \ref{sec:conclusion}, we discuss potential future work to improve stableKanren further.

\section{Preliminaries}
    In this section, we review a few definitions, including definite programs (Definition \ref{def:definite-program}), normal programs (Definition \ref{def:normal-program}), stable model semantics (Definition \ref{def:stable-model}), and a new paradigm: answer set programming (ASP).
    We show an equivalent example of Listing \ref{lst:sk_comb} written in ASP (Listing \ref{lst:normal}) producing multiple models.
    Then, we review the definition of integrity constraints (Definition \ref{def:constraint-rule}) in ASP and the usage of integrity constraints to prune unwanted solutions (Listing \ref{lst:cons}).
    Lastly, we compare and contrast two existing approaches to handling integrity constraints and explain why we need a new approach to extend stableKanren.

\subsection{Integrity Constraints}
\label{sec:predicate-constraints}
    Let us define integrity constraints in answer set programming (ASP).
    To begin with, we use the definition of \textit{definite program} and \textit{normal program} from John W. Lloyd \cite{DBLP:books/sp/Lloyd87}.

    \begin{definition}[definite program clause]
    A \textit{definite program clause} is a clause of the form,
    \[A \leftarrow B_1, \cdots, B_n\]
    where $A, B_1, \dots , B_n$ are atoms.
    \end{definition}
    A definite program clause contains precisely one atom A in its consequent.
    $A$ is called the \textit{head} and $B_1, \dots , B_n$ is called the \textit{body} of the program clause.
    \begin{definition}[definite program]
    \label{def:definite-program}
    A \textit{definite program} is a finite set of definite program clauses.
    \end{definition}
    
    Based on the definite program clause's definition, we have the definition for \textit{normal program clause} and \textit{normal program}.
    \begin{definition}[normal program clause]
    \label{def:normal-program-clause}
    A \textit{normal program clause} is a clause of the form,
    \[A \leftarrow B_1, \cdots , B_n, not \; B_{n+1}, \cdots , not \; B_{m}\]
    \end{definition}
    For a normal program clause, the body of a program clause is a conjunction of literals instead of atoms, $B_1, \cdots, B_n$ are \textit{positive literals} and $not \; B_{n+1}, \cdots, not \; B_{m}$ are \textit{negative literals}.
    \begin{definition}[normal program]
    \label{def:normal-program}
    A \textit{normal program} is a finite set of normal program clauses.
    \end{definition}
    The semantics of the definite and normal programs are minimal model semantics \cite{Emden/Kowalski:1976:Semantics-of-Positive-LP} and stable model semantics \cite{Gelfond:1988:stable}, respectively.

\subsubsection{Stable Model Semantics}
\label{sec:stable-model-semantics}
    Michael Gelfond and Vladimir Lifschitz introduce \textit{stable model semantics} as the semantics for normal programs \cite{Gelfond:1988:stable}.
    Later, Miroslaw Truszczynski proposed an alternative reduct to the original definition: the alternative reduct leaves clause heads intact and only reduces clause bodies \cite{truszczynski2012connecting}.
    \begin{definition}[stable model semantics]
    \label{def:stable-model}
    Given an input program $P$, the first step is getting a \emph{propositional image} of $P$.
    A propositional image $\Pi$ is obtained from grounding each variable in $P$.
    The second step is enumerating all interpretations $I$ of $\Pi$.
    For a $\Pi$ that has $N$ atoms, we will have $2^N$ interpretations.
    The third step is using each model $M$ from $I$ to create a \emph{reduct program} $\Pi_M$ and verify $M$ is the minimal model of $\Pi_M$.
    To create a reduct program, we replace a negative literal $\lnot B_i$ in the clause with $\bot$ if $B_i \in M$; otherwise, we replace it with $\top$.
    Once completed, $\Pi_M$ is negation-free and has a unique minimal model $M'$.
    If $M = M'$, we say $M$ is a stable model of $P$.
    \end{definition}

    Stable model semantics handles negative literals in a normal program by using reduct programs and the minimal model of reduct programs.
    Unlike minimal model semantics, which has only one model for a definite program, stable model semantics can have three outcomes: no model, one model, or multiple models for a normal program.
    Generally speaking, negative literals allow the user to express contradictions in logic programs, and the stable model semantics resolve contradictions to produce results.
    If we cannot resolve contradictions in the problem, we have no models or solutions.
    If we can find more than one way to resolve contradictions in the problem, we have multiple models or solutions.

\subsubsection{Answer Set Programming}
\label{sec:asp}
    As we can see, negative literals in normal programs grant more power to logic programming, and the stable model semantics provide proper interpretation for normal programs.
    A new paradigm named answer set programming (ASP) emerged \cite{Marek1999}.
    ASP extends Prolog by combining negative literals with other expressive constructs like choice rules, integrity constraints, and aggregators so the user can encode declarative solutions to combinatorial search problems.
    For example, given an ASP program in Listing \ref{lst:normal},
    \begin{lstlisting}[caption=An ASP program produces multiple models, label={lst:normal}]
pick(X, Y) :- num(X), num(Y), not free(X, Y).
free(X, Y) :- num(X), num(Y), not pick(X, Y).
num(X) :- X = 1..3.
    \end{lstlisting}
    The above program describes a three-by-three board of nine positions in total.
    Each position has either pick or not pick choices.
    So, it produces all possible combinations, $2^9 = 512$ models, under stable model semantics.

    The term ``constraint'' used in ASP differs from that used in constraint logic programming (CLP).
    The constraint in CLP focuses on the domain of variables, but the constraint in ASP focuses on the truth value of predicates.
    The \textit{integrity constraint} in ASP is defined as a headless normal program clause.
    \begin{definition}[integrity constraint]
    \label{def:constraint-rule}
    A \textit{integrity constraint} is a clause of the form,
    \[ \bot \leftarrow B_1, \cdots , B_n, not \; B_{n+1}, \cdots , not \; B_{m}\]
    \end{definition}
    For an integrity constraint, the head of a clause is $\bot$ (false) instead of an atom.
    The constraint is violated only if all body literals are evaluated as $\top$ (true).
    When the clause's body is evaluated as $\top$, we get a formula that says true implies false.
    \[\bot \leftarrow \top\]
    This formula is impossible to prove true.
    So, the constraint is violated.
    The constraint can be satisfied if any literal in the body is evaluated to be false.
    Therefore, ASP integrity constraints mean that not all literals in the body can be proven true.

    As we have mentioned in section \ref{sec:stable-model-semantics}, a normal program produces multiple models to resolve the contradictions caused by negative literals.
    Here, integrity constraints control the number of models a program can produce.
    Applying integrity constraints, we can prune out unwanted models.
    For instance, let us add an integrity constraint to our previous example (Listing \ref{lst:normal}) in Listing \ref{lst:cons}.
    \begin{lstlisting}[caption=An ASP program with integrity constraint, label={lst:cons}]
pick(X, Y) :- num(X), num(Y), not free(X, Y).
free(X, Y) :- num(X), num(Y), not pick(X, Y).
num(X) :- X = 1..3.

:- pick(X, Y), pick(U, V), X = U, Y != V.
    \end{lstlisting}
    The integrity constraint on the last line indicates that we do not pick the position in the same row.
    Therefore, the total number of models reduces from 512 to 64.
    If we add one more integrity constraint to prevent the answer from taking the position in the same column, the total number of models decreases to 34.

\subsection{Predicate Constraint Handling Methods}
\label{sec:bottom-up-top-down}
    We review existing integrity constraint handling methods to see if they can extend stableKanren. 
    There are two approaches, bottom-up and top-down, to handling integrity constraints.
    
    The bottom-up approach \cite{kaminski2023foundations, Gebser07conflict-drivenanswer} uses \emph{grounding} and \emph{constraint propagation} to obtain a model of the input program.
    In the bottom-up approach, the integrity constraints are treated the same as other normal clauses, so the grounding removes variables in the integrity constraint.
    For example, the integrity constraint we used to remove the same row in Listing \ref{lst:cons} produces 18 propositional constraints after grounding.
    \begin{lstlisting}
:-pick(1,1),pick(1,2). :-pick(1,1),pick(1,3). :-pick(2,1),pick(2,2).
:-pick(2,1),pick(2,3). :-pick(3,1),pick(3,2). :-pick(3,1),pick(3,3).
:-pick(1,2),pick(1,1). :-pick(1,2),pick(1,3). :-pick(2,2),pick(2,1).
:-pick(2,2),pick(2,3). :-pick(3,2),pick(3,1). :-pick(3,2),pick(3,3).
:-pick(1,3),pick(1,1). :-pick(1,3),pick(1,2). :-pick(2,3),pick(2,1).
:-pick(2,3),pick(2,2). :-pick(3,3),pick(3,1). :-pick(3,3),pick(3,2).
    \end{lstlisting}
    At the runtime, each grounded atom assigns a truth value; if any rules are violated, the solver adjusts the truth assignment.
    However, the grounding stage leaves a heavy memory footprint.
    There may have been unused propositional constraints during solving, but the solver still generates them.
    Also, it keeps all constraints even if they are satisfied.
    This memory bottleneck becomes a big issue when applying the bottom-up solver to a large-scale problem.
    Lastly, it does not align with the top-down nature of stableKanren.
    
    The top-down approach \cite{Marple:2017:SASP, DBLP:journals/corr/abs-1804-11162} introduces a head to the headless integrity constraints and appends these transformed constraints at the end of resolution.
    The integrity constraints must add a \textit{false} to the heads.
    Internally, it uses \textit{NMR\_CHECK transformation} \cite{Marple:2017:SASP} to convert each integrity constraint into nested \textit{forall} so that not all constraints are generated at the beginning but after resolution.
    For example, an equivalent of Listing \ref{lst:cons} is written in s(CASP) syntax in Listing \ref{lst:gc-scasp}.
    \begin{lstlisting} [caption=An ASP program with integrity constraint in s(CASP), label=lst:gc-scasp]
:- use_module(library(scasp)).
num(1). num(2). num(3).
pick(X, Y) :- num(X), num(Y), not free(X, Y).
free(X, Y) :- num(X), num(Y), not pick(X, Y).
false :-  pick(X, Y1), pick(X, Y2), Y1 \= Y2.
    \end{lstlisting}
    However, resolution keeps solving these appended constraints instead of verifying them, which causes them not to work as expected.
    A query on the above program using s(CASP) \footnote{\url{https://swish.swi-prolog.org/example/scasp.swinb}},
        \begin{lstlisting}
? pick(1, 1), pick(1, 2).
    \end{lstlisting}
    produces a wrong answer, saying $(1, 1)$ and $(1, 2)$ can be picked at the same time, but the constraint does not allow this combination.
    Hence, the top-down approach could not properly handle ASP integrity constraints.
    The issues are their \textit{NMR\_CHECK transformation} on integrity constraints \cite{Marple:2017:SASP} as well as their resolution process.
    Therefore, it cannot be used to extend stableKanren.

\section{Extending stableKanren}
\label{sec:our-method}
    The integrity constraint in Definition \ref{def:constraint-rule} is a headless clause, but stableKanren requires a valid head to create a goal function.
    As we have shown in Section \ref{sec:bottom-up-top-down}, a suitable top-down approach is needed for handling integrity constraints.
    Therefore, we design our approach to handling integrity constraints.
    Then, we apply our design to extend stableKanren.

    One possible idea is to fix the existing top-down issues.
    We can introduce a head to the headless integrity constraint but in a different way.
    We translate it into a clause that creates a contradiction, as shown in Definition \ref{def:alter-constraint-rule}.
    \begin{definition}[alternative integrity constraint]
    \label{def:alter-constraint-rule}
    An \textit{alternative integrity constraint} is a clause of the form,
    \[ \boldsymbol{fail} \leftarrow B_1, \cdots , B_n, not \; B_{n+1}, \cdots , not \; B_{m}, \boldsymbol{not \; fail}\]
    \end{definition}
    This translation is equivalent to the original integrity constraint in Definition \ref{def:constraint-rule}, and no other transformations are needed.
    The \textit{fail} and \textit{not fail} pair introduces a contradiction when all other clause's bodies $B_i$ are evaluated as true.
    The alternated integrity constraints implicitly append to the end of resolution.
    Then, we also modify resolution to handle these special goals.
    Modifying resolution from proving a goal to checking the proven goals.

    Even though the issues are fixed, the above generate-and-test approach waits until the very last moment to determine if a constraint is violated, even if the constraint can terminate the computation early.
    We want to propagate the constraints as early as possible to save unnecessary computation like what the bottom-up approach achieved.
    However, we also do not want to ground all pairs of constraints initially.

    To develop a new approach, we should leave integrity constraints on the side of resolution as exceptions to track the integrity constraint that could be violated during resolution and terminate the resolution branch early.
    From Section \ref{sec:asp}, we know that in top-down solving, resolution always proves a goal (predicate), either positive or negative, to be true, and the integrity constraints prevent a set of goals or predicates from being true.
    Integrity constraints behave like exceptions in resolution, where the regular resolution computation does not always trigger a constraint violation, and the constraint handler terminates the resolution once a violation is raised.
    We aim to use integrity constraints to create a handler for the top-down resolution.
    
\subsection{Design Overview}
\label{sec:design-idea}
    To treat integrity constraints as exceptions during resolution, we categorize the bodies ($B_i$) in an integrity constraint (Definition \ref{def:constraint-rule}) into two types: an \textit{emitter} and a \textit{verifier}.
    \begin{definition}[emitter]
    \label{def:emitter}
        An \textit{emitter} is a predicate (goal function) that emits values.
    \end{definition}
    \begin{definition}[verifier]
    \label{def:verifier}
        A \textit{verifier} is a boolean expression that receives and verifies values from the emitter.
    \end{definition}
    All verifiers from the same integrity constraint are combined in a constraint handler.
    \begin{definition}[constraint handler]
    \label{def:constraint-handler}
        A constraint handler is a concatenation (\textit{and} operator) of verifiers and $\top$ (true).
    \end{definition}
    Therefore, if all verifiers are true, the constraint handler returns true, indicating that a constraint has been violated.
    Like turning a traditional list into a lazy delayed list utilized continuation, breaking an integrity constraint into two parts allows us to introduce continuation.
    In our case, the verifier is the continuation of the emitter.
    Integrity constraints control the predicate (goal function) outcome resolution produces, and the emitters are implanted as checkpoints in resolution to emit values to verifiers.
    So, the verifiers in the constraint handler are waiting for the emitter's values to continue checking.
    The variables connecting the emitter and verifier have to be unique.
    When resolution successfully unifies variables with values for a predicate, the values are emitted to the verifiers.

    Integrity constraints control the outcome of the predicates (goal functions), and each predicate corresponds to an emitter.
    So, we have three scenarios depending on the number of emitters: no emitter, one emitter, and two or more emitters.
    Let us use Listing \ref{lst:normal} as an example to show different scenarios of integrity constraints.
    During resolution, three predicates (\textit{num}, \textit{pick}, and \textit{free}) can emit values.
    \begin{lstlisting}
pick(X, Y) :- num(X), num(Y), not free(X, Y).
free(X, Y) :- num(X), num(Y), not pick(X, Y).
num(X) :- X = 1..3.
    \end{lstlisting}
    When there is no emitter in the integrity constraint, the verifiers directly control resolution.
    For example, consider the following integrity constraint.
    \begin{lstlisting}
:- 1 = 1.
    \end{lstlisting}
    It has no emitter but only one verifier ``$1 = 1$'', so the constraint handler is ``$(1 = 1) \land \top$'', and resolution can immediately verify the constraint.
    In this example, the constraint handler evaluates the outcome as true and terminates the resolution.
    The verifiers cannot have any variables as they do not receive any values from the emitter.
    For example, consider the following integrity constraint.
    \begin{lstlisting}
:- X = 1.
    \end{lstlisting}
    There is no way to evaluate the expression $X = 1$ to be true or false.

    When one emitter is in the integrity constraint, the verifiers control resolution based on the emitter's value.
    For example, consider the following integrity constraint.
    \begin{lstlisting}
:- not free(X, Y).
    \end{lstlisting}
    It has one emitter \textit{not free}, but no verifier, so the constraint handler is ``$\top$'', whatever value emitted from \textit{not free} terminates resolution.
    So, \textit{not free} cannot be proven true for any values.
    If we add a verifier to the previous example, it is as follows.
    \begin{lstlisting}
:- not free(X, Y), X = 1.
    \end{lstlisting} 
    Now, it has one emitter \textit{not free}, and one verifier ``$X = 1$'', so the constraint handler is ``$(X = 1) \land \top$''.
    Here, variable $X$ connects the emitter and verifier.
    In this case, if the emitter emits 1 to the verifier, the constraint handler terminates resolution.
    So, \textit{not free} cannot be proven true for a value of 1.
    Consider another integrity constraint as follows.
    \begin{lstlisting}
:- not free(X, Y), X = Y.
    \end{lstlisting}
    In this case, if the emitter emits $X$ and $Y$ to the verifier, and when $X = Y$, the constraint handler terminates resolution.

    When two or more emitters exist in the integrity constraint, the verifiers collect all values before checking the constraint.
    For example, consider the following integrity constraint.
    \begin{lstlisting}
:- pick(X,Y), pick(U,V), X = U, Y != V.
    \end{lstlisting}
    This integrity constraint specifies that the same row should not be picked.
    When the first emitter emits ``$X = 1, Y = 1$'' during resolution, the verifier only receives partial values, the constraint handler updated as ``$(1 = U) \land (1 \neq V) \land \top$'', so the constraint cannot be verified, and resolution continues.
    When the second emitter emits ``$U = 1, V = 2$'' from a later resolution, the verifier receives all values, and the constraint handler updated as ``$(1 = 1) \land (1 \neq 2) \land \top$''.
    The constraint handler was evaluated as true.
    So, the resolution branch stops here.
    
    Notice that predicate \textit{pick} can be both the first and second emitter during resolution.
    If \textit{pick} emits values ``$U = 2, V = 1$'' that did not violate the partial constraint handler ``$(1 = U) \land (1 \neq V) \land \top$'', these values will also be received by a new set of verifiers as first emitter to create another partial constraint handler ``$(2 = U) \land (1 \neq V) \land \top$''.
    The resolution carries on with these partial constraint handlers.
    As resolution continues, more and more partial constraint handlers build up.

\subsection{Constraint Compilation}
\label{sec:constraint-compilation}
    We extend stableKanren to support integrity constraints.
    Our implementation strategy is compiling constraints into emitters and verifiers, implanting emitters on the resolution tree to emit values to verifiers during resolution, and checking verification results.
    This section focuses on constraint compilation, and we will illustrate constraint verification during resolution in Section \ref{sec:constraint-checking}.

    At compilation time, we add a new syntax macro \textit{constrainto} to represent an integrity constraint as emitters and verifiers.
    We follow the assumption in stableKanren about variable safety \cite{stableKanren}.
    We assume the emitters are safe, which means they always emit values to variables.
    This assumption can be relaxed, but we leave it as future work.
    Also, we assume that each emitter uses a unique variable to connect with the verifier, as the last example we described in Section \ref{sec:design-idea}.
    For example, an integrity constraint,
    \begin{lstlisting}
:- pick(X,Y), pick(X,V), Y != V.
    \end{lstlisting}
    needs to be converted as
    \begin{lstlisting}
:- pick(X,Y), pick(U,V), X = U, Y != V.
    \end{lstlisting}
    before using \textit{constrainto}.
    So, \textit{constrainto} takes a list of emitters and a list of verifiers in Listing \ref{lst:constrainto}.
    \begin{lstlisting}[caption=An example constrainto,label=lst:constrainto]
(constrainto [(pick X Y) (pick U V)] [(= X U) (not (= Y V))])
    \end{lstlisting}
    A preprocessing compiler that implicitly breaks a constraint and creates the intermediate variables left as future work.

    As we discussed in Section \ref{sec:design-idea}, an emitter is a valid goal function, and a verifier is a boolean expression.
    Resolution nodes form a resolution tree, each corresponding to a stableKanren execution unit.
    
\begin{figure}[h]
    \centering
    \includegraphics[scale=0.5]{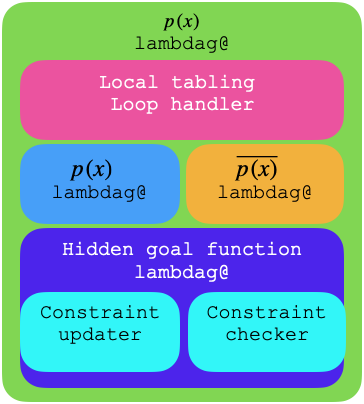}
    \caption{Adding checkpoints to stableKanren execution unit}
    \label{fig:skc-unit}
    \Description{}
    \end{figure}

    In stableKanren \cite{stableKanren}, they use \textit{complement} to implicitly create a negative goal from the positive goal.
    Then, positive and negative goals are wrapped under one unified execution unit through \textit{defineo}.
    Lastly, a hidden goal function for updating the local table is attached at the end of the positive and negative goals.
    Therefore, we modify the hidden goal function to add two checkpoints, \textit{constraint-updater} and \textit{constraint-checker}. 
    The overall changes are illustrated in Figure \ref{fig:skc-unit}.

    The emitters are implanted according to a positive or negative type.
    If the emitter is ``\textit{(pick X Y)}'', then it implants on the internal positive goal $pick(X, Y)$.
    If the emitter is ``\textit{(noto (pick X Y))}'', then it implants on the internal negative goal $\overline{pick(X, Y)}$.
    Verifiers are concatenated to a constraint handler using $and$.
    We use the verifiers from the example in Listing \ref{lst:constrainto} to show a constraint handler.
    \begin{lstlisting}
(and (= X U) (not (= Y V)))
    \end{lstlisting}
    The above constraint handler is a continuation.
    The verifiers are waiting for values from the emitters to continue computation.

\subsection{Constraint Verification}
\label{sec:constraint-checking}
    During resolution, a \textit{constraint-updater} and a \textit{constraint-checker} are used to update and check a constraint handler's verifiers.
    To pass values to a constraint handler (continuation), we wrap anonymous lambdas around it, like in Listing \ref{lst:cons_expr_lambda}.
    \begin{lstlisting}[caption=The constraint handler gets two pairs of values,label=lst:cons_expr_lambda]
((lambda (U V)
   ((lambda (X Y)
      (and (= X U) (not (= Y V)))) 
    1 3))
 2 4)
    \end{lstlisting}
    We define a new macro in Listing \ref{lst:constraint-constructor} for wrapping so that we can wrap arbitrary numbers of layers as needed. 
    \begin{lstlisting}[caption=A macro to wrap a layer of anonymous lambda to a constraint handler,label=lst:constraint-constructor]
(define-syntax constraint-constructor
  (syntax-rules ()
    [(_ (params ...) (values ...) expr)
       `((lambda (params ...)
           expr) values ...)]))
    \end{lstlisting}
    The \textit{constraint-constructor} uses an anonymous lambda to wrap around an expression.
    We cascade apply \textit{constraint-constructor} twice as follows.
    \begin{lstlisting}
(constraint-constructor (U V) (2 4) 
  ,(constraint-constructor (X Y) (1 3)
     (and (= X U) (not (= Y V)))))
    \end{lstlisting}
    We get an expression in the same format as Listing \ref{lst:cons_expr_lambda} with values.

    Every time the emitter generates new values, \textit{constraint-updater} passes the values to the verifiers inside a constraint handler using \textit{constraint-constructor}.
    We use the same constraint handler as an example.
    \begin{lstlisting}
(and (= X U) (not (= Y V)))
    \end{lstlisting}
    When the first emitter emits values $X = 1, Y = 1$, \textit{constraint-updater} wraps one layer around the constraint handler.
    \begin{lstlisting}
(constraint-constructor (X Y) (1 1)
  (and (= X U) (not (= Y V))))
    \end{lstlisting}
    We get a partial constraint handler as follows.
    \begin{lstlisting}
((lambda (X Y)
   (and (= X U) (not (= Y V))))
 1 1)
    \end{lstlisting}
    It cannot use \textit{constraint-checker} to check or get a result as it is missing two values.
    Then, the second emitter emits values $U = 1, V = 2$, \textit{constraint-updater} wraps one more layer around the partial constraint handler.
    \begin{lstlisting}
(constraint-constructor (U V) (1 2)
  ((lambda (X Y)
   (and (= X U) (not (= Y V))))
 1 1))
    \end{lstlisting}
    We get a complete constraint handler as follows.
    \begin{lstlisting}
((lambda (U V)
   ((lambda (X Y) 
      (and (= X U) (not (= Y V))))
    1 1))
 1 2)
    \end{lstlisting}
    Now, it is ready to use \textit{constraint-checker} to check and get a constraint violation to terminate the resolution branch.
    If the second emitter emits a different value $U = 2, V = 1$, \textit{constraint-updater} wraps another layer around the partial constraint handler.
    \begin{lstlisting}
(constraint-constructor (U V) (2 1)
  ((lambda (X Y)
   (and (= X U) (not (= Y V))))
 1 1))
    \end{lstlisting}
    We get a different complete constraint handler, which is as follows,
    \begin{lstlisting}
((lambda (U V)
   ((lambda (X Y)
      (and (= X U) (not (= Y V))))
    1 1))
 2 1)
    \end{lstlisting}
    This time, \textit{constraint-checker} did not find a violated constraint, and the resolution continues.
    Recall the discussion we had at the end of Section \ref{sec:design-idea}.
    The verified complete constraint handler is discarded, but resolution carries the partial constraint handlers for future checking.
    The number of partial constraint handlers (continuations) is gradually growing, so the search space is narrowing as resolution continues.

\section{Using Extended stableKanren}
\label{sec:result}
    This section demonstrates a general approach to solving combinatorial problems using integrity constraints in extended stableKanren.
    The general steps for solving these problems are: firstly, modeling the problem as a graph; secondly, generating all combinations of the nodes or edges that are picked or not picked; and lastly, pruning unwanted solutions using constraints derived from the desired solution's property.
    We use three examples, the nqueens problem, graph coloring, and the Hamiltonian cycle, to apply the general steps.
    The key idea is not to describe ``how'' to solve a problem but to focus on ``what'' is the solution's property.
    We show that the same solution can run with different queries for various purposes. 
\subsection{The Nqueens Problem}
\label{sec:nqueens}
    Let us try to solve the nqueens problem using a new approach.
    The nqueens problem is given $n$ queens to place them on a $n \times n$ chessboard so that no two queens can attack one another.

    Firstly, the graph representation of the problem is the $n \times n$ chessboard, and a stableKanren program can represent this graph.
    For example, a goal function \textit{num} represents $4$ numbers of a $4 \times 4$ board.
    \begin{lstlisting}
(defineo (num x) (conde [(== x 1)] [(== x 2)] [(== x 3)] [(== x 4)]))
    \end{lstlisting}
    The above is the problem instance part, where it depends on the actual problem description.
    Secondly, a queen can be placed or not placed in each position on the board.
    So there are $2^{n \times n}$ combinations of the queen placements.
    Lastly, unwanted queen placements are pruned out to get answers.
    These steps are written in stableKanren in Listing \ref{lst:nq-constraint}.
    \begin{lstlisting}[caption=Nqueens in stableKanren with integrity constraints, label=lst:nq-constraint]
(defineo (queen x y) (num x) (num y) (noto (free x y)))
(defineo (free x y) (num x) (num y) (noto (queen x y)))
(constrainto [(queen x y) (queen u v)] [(= x u) (not (= y v))])
(constrainto [(queen x y) (queen u v)] [(= y v) (not (= x u))])
(constrainto [(queen x y) (queen u v)] [(= (abs (- x u)) (abs (- y v)))
                                           (not (= x u)) (not (= y v))])

(defineo (row x) (fresh (y) (num x) (num y) (queen x y)))
(defineo (col y) (fresh (x) (num x) (num y) (queen x y)))

(constrainto [(num x) (noto (row u))] [(= x u)])
(constrainto [(num y) (noto (col v))] [(= y v)])

(constrainto [(queen x y) (queen u v)] [(> x u)])
    \end{lstlisting}
    The \textit{queen} and \textit{free} generate all combinations of queen placement, either to place or not to place a queen in the position.
    The restriction between two queens is that they cannot be in the same row, column, or diagonal.
    So, three constraints prune out unwanted answers.
    Picking the number of queens fewer than $n$ also satisfied the constraints but not the expected answers.
    The \textit{row} and \textit{col} guarantee each row and column has a queen to ensure that exact $n$ queens are placed on the board.
    The last constraint is top-down query optimization, requiring the queen in ascending order on row numbering.

    A set of queries and outputs for various purposes are shown as follows. 
    \begin{lstlisting}
> (run* (q) (fresh (x1 x2 x3 x4 y1 y2 y3 y4)
              (queen x1 y1) (queen x2 y2) (queen x3 y3) (queen x4 y4)
              (== q `((,x1 ,y1) (,x2 ,y2) (,x3 ,y3) (,x4 ,y4)))))
(*\textbf{(((1 2) (2 4) (3 1) (4 3)) ((1 3) (2 1) (3 4) (4 2)))}*)
> (run* (q) (queen 1 2) (queen 2 4) (queen 3 1) (queen 4 3))
(*\textbf{(\_.0)}*)
> (run* (q) (fresh (x1 x3 x4 y1 y3 y4)
              (queen x1 y1) (queen 2 1) (queen x3 y3) (queen x4 y4)
              (== q `((,x1 ,y1) (,x3 ,y3) (,x4 ,y4)))))
(*\textbf{(((1 3) (3 4) (4 2)))}*)
> (run* (q) (fresh (x1 x3 x4 y1 y3 y4)
              (queen x1 y1) (queen 2 2) (queen x3 y3) (queen x4 y4)
              (== q `((,x1 ,y1) (,x3 ,y3) (,x4 ,y4)))))
(*\textbf{()}*)
    \end{lstlisting}
    The first query is a producer, which generates all solutions.
    It finds two solutions.
    The second query is a checker, which checks whether the given solution is correct or not.
    It says the solution is correct.
    The third query is a prover, which completes the blanks of a partial solution.
    The last query is unable to prove there is an answer given a queen placed at position $(2, 2)$.

\subsection{Graph Coloring}
\label{sec:gc}
\begin{figure}[h]
    \centering
    \includegraphics[scale=0.5]{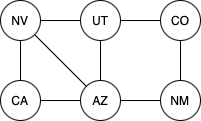}
    \caption{A simple connected graph for coloring}
    \label{fig:states}
    \Description{A simple connected graph with six nodes, representing six southern-west states in the U.S.A.}
    \end{figure}

    Now, let us try to solve the graph coloring problem in extended stableKanren.
    The graph coloring problem is given a connected graph to generate a coloring in which no adjacency nodes use the same color.
    For example, given a graph in Figure \ref{fig:states} and three colors, red, green, and blue, the task is to color it.

    The graph represents the bordering relationship of six states in the southwest region of the United States.
    Firstly, a stableKanren program represents the problem instance as follows.
    \begin{lstlisting}
(defineo (node x)
  (conde [(== x 'AZ)][(== x 'CA)][(== x 'CO)]
         [(== x 'NV)][(== x 'NM)][(== x 'UT)]))

(defineo (edge x y)
  (conde
    [(== x 'AZ) (== y 'CA)][(== x 'AZ) (== y 'NM)][(== x 'AZ) (== y 'NV)]
    [(== x 'AZ) (== y 'UT)][(== x 'CA) (== y 'NV)][(== x 'CO) (== y 'NM)]
    [(== x 'CO) (== y 'UT)][(== x 'NV) (== y 'UT)]))

(defineo (neighbors x y) (conde [(edge x y)] [(edge y x)]))

(defineo (color c) (conde [(== c 'red)] [(== c 'green)] [(== c 'blue)]))
    \end{lstlisting}

    Secondly, each state on the graph can be colored or not colored using the coloring options.
    So there are $2^{6 \times 3}$ combinations of the colorings.
    Lastly, unwanted colorings are pruned out to get answers.
    These two steps are written in stableKanren in Listing \ref{lst:gc-constraint}.
    \begin{lstlisting}[caption=Graph coloring in stableKanren with integrity constraints,label=lst:gc-constraint]
(defineo (assign n c) (node n) (color c) (noto (free n c)))
(defineo (free n c) (node n) (color c) (noto (assign n c)))

(constrainto 
    [(assign n1 c1) (assign n2 c2)]
    [(eq? n1 n2)])
(constrainto
    [(neighbors x y) (assign n1 c1) (assign n2 c2)]
    [(eq? x n1) (eq? y n2) (eq? c1 c2)])

(defineo (assigned n) (fresh (c) (node n) (color c) (assign n c)))
(constrainto [(node n) (noto (assigned m))] [(eq? n m)])

(constrainto 
    [(assign n1 c1) (assign n2 c2)]
    [(> (symbol-hash n1) (symbol-hash n2))])
    \end{lstlisting}
    The \textit{assign} and \textit{free} generate all combinations of colorings, either to color or not to color a state.
    There are three properties of invalid coloring: first, the same state colored more than once; second, the bordering states colored with the same color; third, the state did not get any color.
    Therefore, three constraints prune out unwanted colorings.
    The last constraint is also for top-down query optimization: it requires the node to be in ascending order of hash value.

    A set of queries and outputs are shown as follows.
    \begin{lstlisting}
> (run 1 (q) (fresh (n1 n2 n3 n4 n5 n6 c1 c2 c3 c4 c5 c6)
               (assign n1 c1) (assign n2 c2) (assign n3 c3)
               (assign n4 c4) (assign n5 c5) (assign n6 c6)
               (== q `((,n1 ,c1) (,n2 ,c2) (,n3 ,c3)
                       (,n4 ,c4) (,n5 ,c5) (,n6 ,c6)))))
(*\textbf{(((CA red) (CO green) (AZ green) (NM red) (NV blue) (UT red)))}*)
> (run* (q) (assign 'CA 'green) (assign 'CO 'red) (assign 'AZ 'red)
            (assign 'NM 'blue) (assign 'NV 'blue) (assign 'UT 'green))
(*\textbf{(\_.0)}*)
> (run* (q) (fresh (n1 n4 n5 n6 c1 c4 c5 c6)
              (assign n1 c1) (assign 'CO 'blue) (assign 'AZ 'red)
              (assign n4 c4) (assign n5 c5) (assign n6 c6)
              (== q `((,n1 ,c1) (,n4 ,c4) (,n5 ,c5) (,n6 ,c6)))))
(*\textbf{(((CA green) (NM green) (NV blue) (UT green)))}*)
> (run* (q) (fresh (n1 n3 n5 n6 c1 c3 c5 c6)
              (assign n1 c1) (assign 'CO 'blue) (assign n3 c3)
              (assign 'NM 'blue) (assign n5 c5) (assign n6 c6)
              (== q `((,n1 ,c1) (,n3 ,c3) (,n5 ,c5) (,n6 ,c6)))))
(*\textbf{()}*)
    \end{lstlisting}
    The first query generates a solution to color the map.
    The second query validates a solution.
    The third query fills in the blank of a partially colored map.
    The last query is unable to fill in the blank on a map that Colorado is colored blue, and New Mexico is also colored blue.

\subsection{Hamiltonian Cycle}
\label{sec:hc}
\begin{figure}[h]
    \centering
    \includegraphics[scale=0.15]{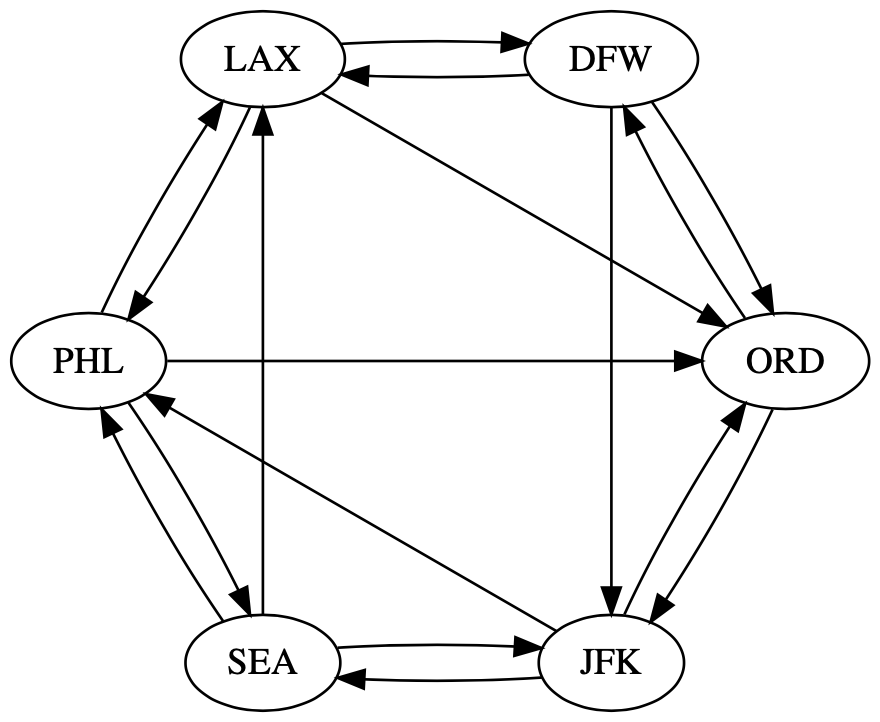}
    \caption{An airline service map}
    \label{fig:airports}
    \Description{An airline service map with six airports, DFW, JFK, LAX, ORD, PHL, and SEA.}
    \end{figure}

    A Hamiltonian cycle is a cycle that visits each vertex exactly once.
    For example, give an airline service map in Figure \ref{fig:airports}, finding a travel plan that visits each airport exactly once.

    The service map is a directed graph, where an arrow means a flight service from one airport to another. Firstly, the problem instance is represented in stableKanren as follows.
    \begin{lstlisting}
(defineo (airport x)
  (conde [(== x 'DFW)] [(== x 'JFK)] [(== x 'LAX)]
         [(== x 'ORD)] [(== x 'PHL)] [(== x 'SEA)]))

(defineo (fly x y)
  (conde [(== x 'DFW) (== y 'JFK)] [(== x 'DFW) (== y 'LAX)]
         [(== x 'DFW) (== y 'ORD)] [(== x 'JFK) (== y 'ORD)]
         [(== x 'JFK) (== y 'PHL)] [(== x 'JFK) (== y 'SEA)]
         [(== x 'LAX) (== y 'DFW)] [(== x 'LAX) (== y 'ORD)]
         [(== x 'LAX) (== y 'PHL)] [(== x 'ORD) (== y 'DFW)]
         [(== x 'ORD) (== y 'JFK)] [(== x 'PHL) (== y 'LAX)]
         [(== x 'PHL) (== y 'ORD)] [(== x 'PHL) (== y 'SEA)]
         [(== x 'SEA) (== y 'JFK)] [(== x 'SEA) (== y 'LAX)]
         [(== x 'SEA) (== y 'PHL)]))
    \end{lstlisting}
    Secondly, each flight service between the two airports can be bought or not bought.
    There are 17 flight services, so it has $2^{17}$ combinations of purchases.
    Lastly, unreasonable purchases are pruned out to get valid travel plans.
    These two steps are written in stableKanren in Listing \ref{lst:hc-constraint}.
    \begin{lstlisting}[caption=Hamiltonian cycle in stableKanren with integrity constraints,label=lst:hc-constraint]
(defineo (buy u v) (airport u) (airport v) (fly u v) (noto (free u v)))
(defineo (free u v) (airport u) (airport v) (fly u v) (noto (buy u v)))

(constrainto [(buy u v) (buy x y)] [(eq? x u) (not (eq? v y))])
(constrainto [(buy u v) (buy x y)] [(not (eq? x u)) (eq? v y)])

(defineo (reachable v)
  (conde
    [(buy 'DFW v)]
    [(fresh (u) (airport u) (buy u v) (reachable u))]))
(constrainto [(airport u) (noto (reachable v))] [(eq? u v)])
    \end{lstlisting}
    The \textit{buy} and \textit{free} generate all combinations of travel plans, either to buy or not to buy a ticket.
    There are three kinds of invalid travel plans: first, flying out from the same airport more than once; second, flying into the same airport more than once; third, not flying to an airport at all.
    Therefore, three constraints prune out unreasonable travel plans.
    Unlike the other two examples, there is no constraint for query optimization.

    A set of queries and outputs of travel plans are shown as follows.
    \begin{lstlisting}
> (run 1 (q) (fresh (a b c d e f)
               (buy a b) (buy b c) (buy c d)
               (buy d e) (buy e f) (buy f a)
               (== q `(,a ,b ,c ,d ,e ,f ,a))))
(*\textbf{((DFW JFK PHL SEA LAX ORD DFW))}*)
> (run* (q) (buy 'JFK 'PHL) (buy 'PHL 'SEA) (buy 'SEA 'LAX)
            (buy 'LAX 'DFW) (buy 'DFW 'ORD) (buy 'ORD 'JFK))
(*\textbf{(\_.0)}*)
> (run* (q) (buy 'JFK 'PHL) (buy 'PHL 'SEA) (buy 'SEA 'DFW)
            (buy 'DFW 'LAX) (buy 'LAX 'ORD) (buy 'ORD 'JFK))
(*\textbf{()}*)
> (run 1 (q) (fresh (a b d e f)
               (buy a b) (buy b 'DFW) (buy 'DFW d)
               (buy d e) (buy e f) (buy f a)
               (== q `(,a ,b DFW ,d ,e ,f ,a))))
(*\textbf{((JFK ORD DFW LAX PHL SEA JFK))}*)
> (run 1 (q) (fresh (a d e f)
               (buy a 'SEA) (buy 'SEA 'DFW) (buy 'DFW d)
               (buy d e) (buy e f) (buy f a)
               (== q `(,a SEA DFW ,d ,e ,f ,a))))
(*\textbf{()}*)
    \end{lstlisting}
    The first query asks for a travel plan.
    It returns one starting at DFW.
    The second query verifies a travel plan starting at JFK.
    It confirms the travel plan is valid.
    The third query verifies another travel plan starting at JFK.
    It says the plan is infeasible.
    The fourth query is about DFW being the third stop during the trip.
    It returns a valid travel plan.
    The last query proposes SEA as the second stop and DFW as the third stop.
    It is impossible to meet the requirement.

\section{Conclusion and Future Work}
\label{sec:conclusion}
    This paper presents a new approach to integrating integrity constraints (Definition \ref{def:constraint-rule}) to stableKanren.
    The challenge is that integrity constraints are headless normal clauses, but a top-down approach like stableKanren requires using the clause's head to create a goal function.
    Our approach turns integrity constraints into exceptions to the top-down resolution.
    We decompose integrity constraints into two parts: emitters (Definition \ref{def:emitter}) and verifiers (Definition \ref{def:verifier}).
    The verifiers can hold the values emitted from the emitters during resolution.
    The extended stableKanren is able to solve combinatorial search problems in a new paradigm.
    The new paradigm focuses on declaring ``what'' are the solutions' properties instead of describing ``how'' to solve a problem.
    The general steps for solving such problems are: first, modeling the problem as a graph; second, generating all combinations of nodes or edges; third, pruning out unwanted combinations.
    We show three examples (Listing \ref{lst:nq-constraint}, \ref{lst:gc-constraint}, and \ref{lst:hc-constraint}) using the new paradigm in extended stableKanren.
    The problem instances are described by using the \textit{defineo} macro.
    The combinations are generated by using the \textit{noto} operator (Listing \ref{lst:sk_comb}).
    The integrity constraints are defined by our new \textit{constrainto} macro (Listing \ref{lst:constrainto}).
    A program written in extended stableKanren can serve as a producer (generating all solutions), a checker (checking a solution), and a prover (filling the blank of a partial solution) without changes to the underlying algorithm.

    There are still many improvements to be made in the future.
    Firstly, we can add more constructs to increase the expressive power.
    The constraints we add in this paper are hard constraints because once a constraint is violated, the partial solution is discarded.
    For some problems, it is acceptable to violate a few constraints with a penalty to create a non-optimal solution.
    Hence, a new top-down soft constraint handling algorithm would be helpful.
    More importantly, as we add declarative constructs, performance optimization becomes critical.
    The user will use these declarative constructs rather than writing their imperative algorithms only if the performance is ideal.
    So, we plan to apply a state-of-the-art conflict-driven nogood learning (CDNL) algorithm to boost our constraint-solving speed \cite{10.5555/1625275.1625336, Gebser07conflict-drivenanswer}.
    Instead of using constraints to check the answer, we can use constraints to guide the search process.
    Porting CDNL to stableKanren solver or other solvers written in a functional language is not simple.
    Chris Hanson and Gerald Jay Sussman proposed a propagation system that contains cells, propagators, dependency-directed backtracking, and can learn the nogoods from contradictions \cite{hanson2021software}.
    Lastly, stableKanren assumes the variables are safe, and we followed that assumption in our integrity constraints extension.
    This assumption can be resolved through compilation time rule rewriting and runtime variable constraints.
    Jason Hemann and Daniel Friedman handle the variable constraints in microKanren \cite{microKanrenConstraints}.
    Adding variable constraints also allows us to combine ASP with constraint logic programming (CLP) in stableKanren, resulting in a more powerful language.



\bibliographystyle{ACM-Reference-Format}
\bibliography{ref}


\end{document}